\documentclass{book}
\usepackage[cip,ChapterTOCs]{sunil}
\usepackage{latexsym}
\usepackage{theorem}
\usepackage{graphicx}
\usepackage{graphics}
\usepackage{epsfig}
\usepackage{makeidx}
\usepackage{showidx}

  \newcommand{\msun}{\mbox{${\rm M}_\odot$}}

\def\apgt{\ {\raise-.5ex\hbox{$\buildrel>\over\sim$}}\ }
\def\aplt{\ {\raise-.5ex\hbox{$\buildrel<\over\sim$}}\ }
\def\lt{\ {\raise-.5ex\hbox{$\buildrel>$}}\ }
\def\gt{\ {\raise-.5ex\hbox{$\buildrel<$}}\ }

\def\aap{\ {A\&A}\ }

\def\apj{\ {ApJ}\ }

\def\apjs{\ {ApJS}\ }

\def\araa{\ {ARA\&A}\ }

\def\mnras{\ {MNRAS}\ }
\def\nat{\ {Nature}\ }

\def\pasp{\ {PASP}\ }

\begin{document}
\tableofcontents

\title{Title text goes here}

\chapterauthor{Alfons G. Hoekstra}{Section Computational Science, University of Amsterdam, Kruislaan 403, Amsterdam, The Netherlands,
alfons@science.uva.nl}
\chapterauthor{Simon Portegies Zwart}{Section
Computational Science and Astronomical Institute ``Anton
Pannekoek'', University of Amsterdam, Kruislaan 403, Amsterdam, The
Netherlands,\\spz@science.uva.nl}
\chapterauthor{Marian Bubak}{Section Computational Science, University of Amsterdam, Kruislaan 403, Amsterdam, The Netherlands,
and AGH University of Science and Technology, Krak\'{o}w, Poland,
bubak@science.uva.nl}
\chapterauthor{Peter M.A. Sloot}{Section
Computational Science, University of Amsterdam, Kruislaan 403,
Amsterdam, The Netherlands, sloot@science.uva.nl}

\chapter{Towards Distributed Petascale Computing}\label{ch01}

\section{Introduction}\label{ch01sec01}

Recent advances in experimental techniques have opened up new
windows into physical and biological processes on many levels of
detail. The resulting data explosion requires sophisticated
techniques, such as grid computing and collaborative virtual
laboratories, to register, transport, store, manipulate, and share
the data. The complete cascade from the individual components to the
fully integrated multi-science systems crosses many orders of
magnitude in temporal and spatial scales. The challenge is to study
not only the fundamental processes on all these separate scales, but
also their mutual coupling through the scales in the overall
multi-scale system, and the resulting emergent properties. These
complex systems display endless signatures of order, disorder,
self-organization and self-annihilation. Understanding, quantifying
and handling this complexity is one of the biggest scientific
challenges of our time \cite{Barabasi2005}.

In this chapter we will argue that studying such multi-scale
multi-science systems gives rise to inherently hybrid models
containing many different algorithms best serviced by different
types of computing environments (ranging from massively parallel
computers, via large-scale special purpose machines to clusters of
PC's) whose total integrated computing capacity can easily reach the
PFlop/s scale. Such hybrid models, in combination with the by now
inherently distributed nature of the data on which the models `feed'
suggest a distributed computing model, where parts of the
multi-scale multi-science model are executed on the most suitable
computing environment, and/or where the computations are carried out
close to the required data (i.e. bring the computations to the data
instead of the other way around).

Prototypical examples of multi-scale multi-science systems come from
bio-medicine, where we have data from virtually all levels between
`molecule and man' and yet we have no models where we can study
these processes as a whole. The complete cascade from the genome,
proteome, metabolome, physiome to health constitutes multi-scale,
multi-science systems, and crosses many orders of magnitude in
temporal and spatial scales \cite{Finkelstein2004, Sloot2006}.
Studying biological modules, their design principles, and their
mutual interactions, through an interplay between experiments and
modeling and simulations, should lead to an understanding of
biological function and to a prediction of the effects of
perturbations (e.g. genetic mutations or presence of drugs).
\cite{Ventura2006}

A good example of the power of this approach, in combination with
state-of-the-art computing environments, is provided by the study of
the heart physiology, where a true multi-scale simulation, going
from genes, to cardiac cells, to the biomechanics of the whole
organ, is now feasible. \cite{Noble2002} This `from genes to health'
is also the vision of the Physiome project \cite{Hunter2003,
Hunter2006}, and the ViroLab \cite{virolab, Sloot2006b}, where a
multi-scale modeling and simulation of human physiology is the
ultimate goal.
 The wealth of data now available from many years
of clinical, epidemiological research and (medical) informatics,
advances in high-throughput genomics and bioinformatics, coupled
with recent developments in computational modeling and simulation,
provides an excellent position to take the next steps towards
understanding the physiology of the human body across the relevant
$10^9$ range of spatial scales (nm to m) and $10^{15}$  range of
temporal scales, ($\mu$s to human lifetime) and to apply this
understanding to the clinic. \cite{Hunter2003, Ayache2005} Examples
of multi-scale modeling are increasingly emerging (see for example,
\cite{Davies2005, Iribe2006, Kelly2006, Sloot2005}).

In Section \ref{ch01sec02} we will consider the Grid as the obvious
choice for a distributed computing framework, and we will then
explore the potential of computational grids for Petascale computing
in Section \ref{ch01sec03}. Section \ref{ch01sec04} presents the
\emph{Virtual Galaxy} as a typical example of a multi-scale
multi-physics application, requiring distributed Petaflop/s
computational power.

\section{Grid Computing}\label{ch01sec02}
The radical increase in the amount of IT-generated data from
physical, living and social systems brings about new challenges
related to the sheer size of data. It was this data `deluge' that
originally triggered the research into grid computing
\cite{Foster2001, Hey2003}. Grid computing is an emerging computing
model that provides the ability to share data and instruments and to
perform high throughput computing by taking advantage of many
networked computers able to divide process execution across a
distributed infrastructure.

As the Grid is ever more frequently used for collaborative problem
solving in research and science, the real challenge is in the
development of new applications for a new kind of users through
virtual organizations. Existing grid programming models are
discussed in \cite{Lee2003, bal2004}.

Workflow  is a convenient way of distribution of computations across
a grid. A large group of composition languages have been studied for
formal description of workflows \cite{aalst2005} and they are used
for orchestration, instantiation, and execution of workflows
\cite{Ludascher2006}. Collaborative applications are also supported
by problem solving environments which enable users to handle
application complexity with web-accessible portals for sharing
software, data, and other resources \cite{pse2005}. Systematic ways
to building grid applications are provided through object-oriented
and component technology, for instance the Common Component
Architecture which combines the IDL-based distributed framework
concept with requirements of scientific applications \cite{cca2006}.
Some recent experiments with computing across grid boundaries,
workflow composition of Grid services with semantic description, and
 development of collaborative problem solving
environments are reported in \cite{malawski2006, wcf2005, cross}.
These new computational approaches should transparently exploit the
dynamic nature of Grid and virtualization of grid infrastructure.
The challenges are efficient usage of knowledge for automatic
composition of applications \cite{kwfgrid}.

Allen et al. in \cite{Allen2003} distinguish four main types of grid
applications: (1) Community-centric; (2) Data-centric ; (3)
Computation-centric; and (4) Inter-action-centric. Data-centric
applications are, and will continue to be the main driving force
behind the Grid. Community-centric applications are about bringing
people or communities together, as e.g. in the Access Grid, or in
distributed collaborative engineering. Interaction-centric
applications are those that require 'a man in the loop', for
instance in real-time computational steering of simulations or
visualizations (as e.g. demonstrated by the CrossGrid project
\cite{cross}.

In this chapter we focuss on Computation-centric applications. These
are the traditional High Performance Computing (HPC) and High
Throughput Computing (HTC) applications which, according to Allen et
al. \cite{Allen2003} ``turned to parallel computing to overcome the
limitations of a single processor, and many of them will turn to
Grid computing to overcome the limitations of a parallel computer.''
In the case of parameter sweep (i.e. HTC) applications this has
already happened. Several groups have demonstrated successful
parameter sweeps on a computational Grid (see e.g.
\cite{Sudholt2004}). For tightly coupled HPC applications this is
not so clear, as common wisdom is that running a tightly coupled
parallel application in a computational grid (in other words, a
parallel job actually running on several parallel machines that
communicate with each other in a Grid) is of no general use because
of the large overheads that will be induced by communications
between computing elements (see e.g. \cite{Lee2003}). However, in
our opinion this certainly is a viable option, provided the
granularity of the computation is large enough to overcome the
admittedly large communication latencies that exist between compute
elements in a Grid. \cite{Hoekstra2005} For PFlop/s scale computing
we can assume that such required large granularity will be reached.
Recently a Computation-centric application running in parallel on
compute elements located in Poland, Cyprus, Portugal, and the
Netherlands was successfully demonstrated \cite{Tirado2005,
Gualandris2007}.

 \section{Petascale Computing on the Grid}\label{ch01sec03}

Execution of multi-scale multi-science models on computational grids
will in general involve a diversity of computing paradigms. On the
highest level functional decompositions may be performed, splitting
the model in sub-models that may involve different types of physics.
For instance, in a fluid-structure interaction application the
functional decomposition leads to one part modeling the structural
mechanics, and another part modeling the fluid flow. In this example
the models are tightly coupled and exchange detailed information
(typically, boundary conditions at each time step). On a lower level
one may again find a functional decomposition, but at some point one
encounters single-scale, single-physics sub-models, that can be
considered as the basic units of the multi-scale multi-science
model. For instance, in a multi-scale model for crack propagation,
the basic units are continuum mechanics at the macroscale, modeled
with finite elements, and molecular dynamics at the microscale
\cite{Broughton1999}. Another examplex is provided by Plasma
Enhanced Vapor Deposition where mutually coupled chemical, plasma
physical and mechanical models can be distinguished \cite{Lera2005}.
In principle all basic modeling units can be executed on a single
(parallel) computer, but they can also be distributed to several
machines in a computational grid.

These basic model units will be large scale simulations by
themselves. With an overall performance on the PFlop/s scale, it is
clear that the basic units will also be running at impressive
speeds. It is difficult to estimate the number of such basic model
units. In the example of the fluid-structure interaction, there are
two, running concurrently. However, in case of for instance a
multi-scale system modeled with the Heterogeneous Multiscale Method
\cite{E2007} there could be millions of instances of a microscopic
model that in principle can execute concurrently (one on each
macroscopic grid point). So, for the basic model units we will find
anything between single processor execution and massively parallel
computations.

A computational grid offers many options of mapping the computations
to computational resources. First, the basic model units can be
mapped to the most suitable resources. So, a parallel solver may be
mapped to massively parallel computers, whereas for other solvers
special purpose hardware may be available, or just single PC's in a
cluster. Next, a distributed simulation system is required to
orchestrate the execution of the multi-scale multi-science models.

A computational grid is an appropriate environment for running
functionally decomposed distributed applications. A good example of
research and development in this area is the CrossGrid Project which
aimed at elaboration of an unified approach to development and
running large scale interactive distributed, compute- and
data-intensive applications, like biomedical simulation and
visualization for vascular surgical procedures, a flooding crisis
team decision support system, distributed data analysis in high
energy physics, and air pollution combined with weather forecasting
\cite{cross}. The following issues were of key importance in this
research and will also play a pivotal role on the road towards
distributed PFlop/s scale computing on the Grid: porting
applications to the grid environment; development of user
interaction services for interactive startup of applications, online
output control, parameter study in the cascade, and runtime
steering, and on-line, interactive performance analysis based
on-line monitoring of grid applications. The elaborated CrossGrid
architecture consists of a set of self-contained subsystems divided
into layers of applications, software development tools and Grid
services \cite{cro-arch}.

Large scale grid applications require on-line performance analysis.
The application monitoring system, OCM-G, is a unique online
monitoring system in which requests and response events are
generated dynamically and can be toggled at runtime. This imposes
much less overhead on the application and therefore can provide more
accurate measurements for the performance analysis tool like G-PM,
which can  display (in form of various metrics) the behavior of Grid
applications \cite{ocm-g}.

The High Level Architecture (HLA) fulfills many requirements of
distributed interactive applications. HLA and the Grid may
complement each other to support  distributed interactive
simulations. The G-HLAM system supports for execution of legacy HLA
federates on the Grid  without imposing  major modifications of
applications. To achieve efficient execution of HLA-based
simulations on the Grid, we introduced migration and monitoring
mechanisms for such applications. This system has been applied to
run two complex distributed interactive applications: N-body
simulation and virtual bypass surgery \cite{cro-hla}.

In the next section we explore in some detail a prototypical
application where all the aforementioned aspects need to be
addressed to obtain distributed Petascale computing.

\section{The Virtual Galaxy}\label{ch01sec04}
\begin{figure}
\begin{center}
  \psfig{figure=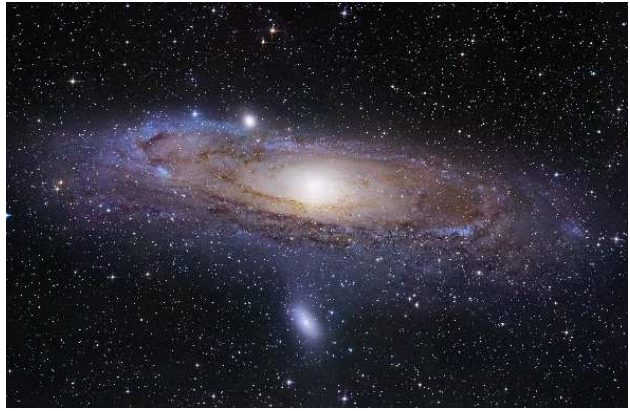,width=0.7\textwidth}
\end{center}
  \caption{\label{fig:M31} The Andromeda Nebula, M31. A mosaic of hundreds of Earth based telescope
pointings were needed to make this image.}
\end {figure}

A grand challenge in computational astrophysics, requiring \emph{at
least} the PFlop/s scale, is the simulation of the physics of
formation and evolution of large spiral galaxies like the Milky-way.
This requires the development of a hybrid simulation environment to
cope with the multiple time scales, the broad range of physics and
the shear number of simulation operations \cite{Makino2005a,
Hut2006}. The nearby grand design spiral galaxy M31 in the
constellation andromeda, as displayed in Fig.\,\ref{fig:M31},
provides an excellent birdseye view of how the Milky-way probably
looks.

This section presents the Virtual Galaxy as a typical example of a
multi-physics application that requires PFlop/s computational
speeds, and has all the right properties to be mapped to distributed
computing resources. We will introduce in some detail the relevant
physics and the expected amount of computations (i.e. Flop) needed
to simulate a Virtual Galaxy. Solving Newton's equations of motion
for any number of stars is a challenge by itself, but to perform
this in an environment with the number of stars as in the Galaxy,
and over the enormous range of density contrasts and with the
inclusion of additional chemical and nuclear physics, doesn't make
the task easier. No single computer will be able to perform the
resulting multitude of computations, and therefore it provides a
excellent example for a hybrid simulation environment containing a
wide variety of distributed hardware. We end this section with a
discussion on how a Virtual Galaxy simulation could be mapped to a
PFlop/s scale grid computing environment. We believe that the
scenarios that we outline are prototypical and also apply to a
multitude of other multi-science multi-scale systems, like the ones
that were discussed section~\ref{ch01sec01} and \ref{ch01sec03}.

\subsection{A Multi-Physics model of the Galaxy}

\begin{figure}
\begin{center}
  \psfig{figure=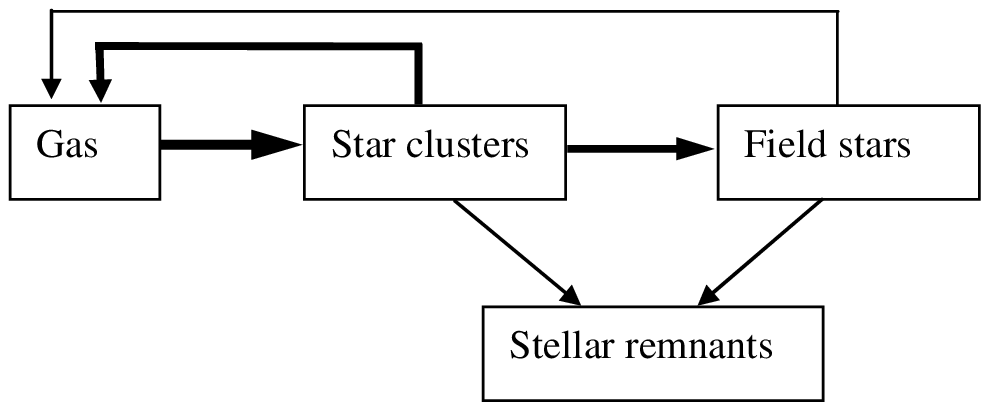,width=0.5\textwidth}
\end{center}
  \caption{\label{fig:gas2gas} Schematic representation of the
  evolution of the gas content of the Galaxy.}
\end{figure}

The Galaxy today contains a few times $10^{11}$ the solar mass
(\msun) in gas and stars. The life cycle of the gas in the Galaxy is
illustrated in Fig.\,\ref{fig:gas2gas}, where we show how gas
transforms to star clusters, which again dissolve to individual
stars. The ingredients for a self consistent model of the Milky-way
Galaxy is based on these same three ingredients: the gas, the star
clusters and the field stellar population. The computational cost
and physical complexity for simulating each of these ingredients can
be estimated based on the adopted algorithms.

\subsubsection{How gas turns into star clusters}
\label{Sect:gas2SCs}\label{Sect:GD}

Stars and star clusters form from giant molecular clouds which
collapse when they become dynamically unstable. The formation of
stars and star clusters is coupled with the galaxy formation
process. The formation of star clusters themselves has been
addressed by many research teams and most of the calculations in
this regard are a technical endeavor which is mainly limited by the
lack of resources.

Simulations of the evolution of a molecular cloud up to the moment
it forms stars are generally performed with adaptive mesh refinement
and smoothed particles hydrodynamics algorithms. These simulations
are complex, and some calculations include turbulent motion of the
gas \cite{Bate2005}, solve the full magnetic hydrodynamic equations
\cite{Zengin2004,Whitehouse2006}, or include radiative transport
\cite{Padoan2002}. All the currently performed dynamical cloud
collapse simulations are computed with a relatively limited accuracy
in the gravitational dynamics. We adopt the smoothed particle
hydrodynamics methodology to calculate the gravitational collapse of
a molecular cloud, as it is relatively simple to implement and has
scalable numerical complexity.  These simulation environments are
generally based on the Barnes-Hut tree code \cite{Barnes1986} for
resolving the self gravity between the gas or dust volume or mass
elements, and have a ${\cal O}(n_{\rm SPH}\log n_{\rm SPH})$ time
complexity \cite{Kawai2004}.

Simulating the collapse of a molecular cloud requires at least $\sim
10^3$ SPH particles per star, a star cluster that eventually (after
the simulation) consisting of ${\cal O}(10^4)$ stars then requires
about $n_{\rm SPH} \sim 10^{7}$ SPH particles.

The collapse of a molecular clound lasts for about $\tau_J \simeq
1/\sqrt{G \rho}$, which for a $10^4$\msun\, molecular cloud with a
size of 10\,pc is about a million years. Within this time span the
molecular cloud will have experienced roughly $10^4$ dynamical time
scales totaling the CPU requirements to about ${\cal O}(10^{11})$
Flop for calculating the gravitational collapse of one molecular
cloud.

\subsubsection{The evolution of the individual stars}
\label{Sect:stars2gas}\label{Sect:SE}

Once most of the gas is cleared from the cluster environment, an
epoch of rather clean dynamical evolution mixed with the evolution
of single stars and binaries starts. In general, star cluster
evolution in this phase may be characterized by a competition
between stellar dynamics and stellar evolution.  Here we focus
mainly on the nuclear evolution of the stars.

With the development of shell based Heny\'e codes
\cite{Eggleton2006} the nuclear evolution of a single star for its
entire lifetime requires about $10^9$ Flop \cite{Makino1990}. Due to
efficient step size refinement the performance of the algorithm is
independent of the lifetime of the star; a 100\,\msun\,star is as
expensive in terms of compute time as a 1\,\msun\, star. Adopting
the mass distribution with which stars are born \cite{Kroupa1990}
about one in 6 stars require a complete evolutionary calculation.
The total compute time for evolving all the stars in the Galaxy over
its full life time then turns out to be about $10^{20}$\,Flop.

Most ($\apgt 99$\%) of all the stars in the Galaxy will not do much
apart from burning their internal fuel.  To reduce the cost of
stellar evolution we can therefore parameterize the evolution of
such stars. Excellent stellar evolution prescriptions at a fraction
of the cost ($\aplt 10^4$ Flop) are available
\cite{Eggleton1989,Hurley2000}, and could be used for the majority
of stars (which is also what we adopted in \S\,\ref{Sect:PM}).

\subsubsection{Dynamical evolution} \label{SCs2stars}\label{Sect:SD}

When a giant molecular cloud collapses one is left with a
conglomeration of bound stars and some residual gas. The latter is
blown away from the cluster by the stellar winds and supernovae of
the young stars. The remaining gas depleted cluster may subsequently
dissolve in the background on a time scale of about $10^8$\,years.

The majority (50-90\%) of star clusters which are formed in the
Galaxy dissolve due to the expulsion of the residual gas
\cite{Goodwin1997, Boily2003}.  Recent reanalysis of the cluster
population of the Large Magelanic cloud indicates that this process
of {\em infant mortality} is independent of the mass of the cluster
\cite{Lamers2005}. Star clusters that survive their infancy engage
in a complicated dynamical evolution which is quite intricately
coupled with the nuclear evolution of the stars
\cite{Portegies2001b}.

The dynamical evolution of a star cluster is best simulated using
direct $N$-body integration techniques, like NBODY4
\cite{Aarseth1975,Aarseth1999} or the {\tt starlab} software
environment \cite{Portegies2001b}.

For dense star clusters the compute time is completely dominated by
the force evaluation. Since each star has a gravitational pull at
all other stars this operation scales with ${\cal O}(N^2)$ for one
dynamical time step. The good news is that the large density
contrast between the cluster central regions and its outskirts can
cover 9 orders of magnitude, and stars far from the cluster center
are regularly moving whereas central stars have less regular orbits
\cite{Gemmeke2006}. By applying smart time stepping algorithms one
can reduce the ${\cal O}(N^2)$ to ${\cal O}(N^{4/3})$ without loss
of accuracy \cite{Makino1988}. In fact one actually gains accuracy
since taking many unnecessary small steps for a regularly integrable
star suffers from numerical round-off.

The GRAPE-6, a special purpose computer for gravitational $N$-body
simulations, performs dynamical evolution simulations at a peak
speed of about 64\,Tflop/s \cite{Makino2001}, and is extremely
suitable for large scale $N$-body simulations.

\subsubsection{The galactic field stars}\label{Sect:FS}

Stars that are liberated by star clusters become part of the
Galactic tidal field. These stars, like the Sun, orbit the Galactic
center in regular orbits. The average time scale for one orbital
revolution for a field star is about 250\,Myr. These regularly
orbiting stars can be resolved dynamically using a relatively
unprecise $N$-body technique, we adopt here the ${\cal O}(N)$
integration algorithm which we introduced in \S\,\ref{Sect:GD}.

In order to resolve a stellar orbit in the Galactic potential about
100 integration time steps are needed.  Per Galactic crossing time
(250\,Myr) this code then requires about $10^6$ operations per star,
resulting in a few times $10^7 N$ Flop for simulating the field
population. Note that simulating the galactic field population is a
trivially parallel operation, as the stars hover around in their
self generated potential

\subsection{A performance model for simulating the
Galaxy}\label{Sect:PM}

Next we describe the required computer resources as a function of
life time of a Virtual Galaxy. The model is relatively simple and
the embedded physics is only approximate, but it will give an
indication on what type of calculation is most relevant in what
state of the evolution of the Galaxy.

According to the model we start the evolution of the Galaxy with
amorphous gas. We subsequently assume that molecular clouds are
formed with power-law mass function with an index of -2 between
$10^3$\,\msun\, and $10^7$\,\msun, with distribution in time which
is flat in $\log t$. We assume that the molecular cloud lives for
between 10\,Myr and 1\,Gyr (with an equal probability between these
moments). The star formation efficiency is 50\%, and the cluster has
an 80\% change to dissolve within 100\,Myr (irrespective of the
cluster mass). The other 20\% clusters dissolve on a time scale of
about $t_{\rm diss} \sim 10 \sqrt{R^3 M}$\,Myr. During this period
they lose mass at a constant rate. The field population is enriched
with the same amount of mass.

\begin{figure}
\begin{center}
  \psfig{figure=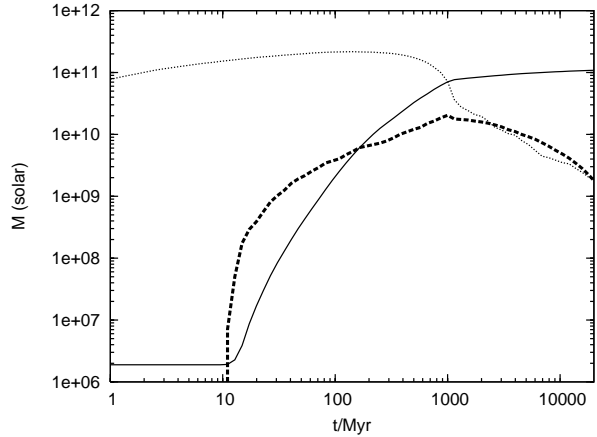,width=0.7\textwidth}
  \end{center}
  \caption{\label{fig:VG} The evolution of the mass content in the
    Galaxy via the simple model described in \S\,\ref{Sect:PM}.  The
    dotted curve give the total mass in giant molecular clouds, the
    thick dashed curve in star clusters and the solid curve in field
    stars, which come from dissolved star clusters.
}
\end{figure}

\begin{figure}
\begin{center}
  \psfig{figure=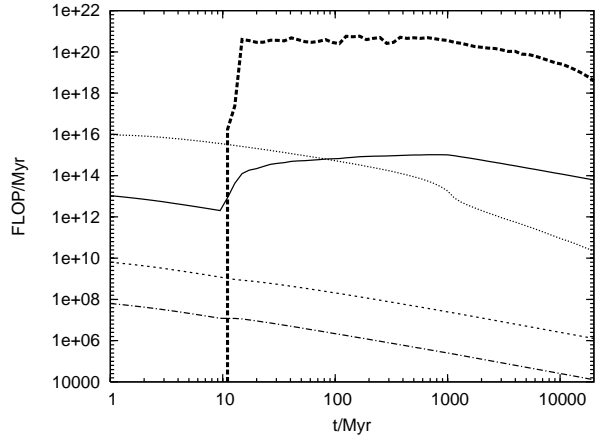,width=0.7\textwidth}
  \end{center}
  \caption{\label{fig:CPU} The number of floating points operations
  expenditure per million years for the various ingredients in the
  performance model.
  The solid, thick short dash and doted curve are as in
  Fig.\,\ref{fig:VG}. New in this figure are the two dotted and
  dash-dotted lines near the bottom, which represent the CPU time
  needed for evolving the field star population (lower dotted curve)
  and dark matter (botton curve). }
\end{figure}

The resulting total mass in molecular clouds, star clusters and
field stars is presented in fig.\,\ref{fig:VG}.  At early age, the
galaxy completely consists of molecular clouds. After about 10\,Myr
some of these cloulds collapse to form star clusters and single
stars, indicated by the rapildy rising solid (field stars) and
dashed (star clusters) curves. The maximum number of star clusters
if reached when the Galaxy is about a Gyr old. The field population
continues to rise to reach a value of a few times $10^{11}$\,\msun\,
at today's age of about 10\,Gyr. By that time the total mass in star
clusters has dropped to several $10^9$\,\msun\, quite comparable
with the observed masses of the field population and the star
cluster content.

In Fig.\,\ref{fig:CPU} we show the evolution of the amount of Flop
required to simulate the entire galaxy, as a function of its life
time. The Flop count along the vertical axis are given in units of
number of floating points operations per million years in Galactic
evolution. For example, to evolve the Galaxy's population of
molecular clouds from 1000\,Myr to 1001\,Myr requires about
$10^{16}$ Flop.

\subsection{Petascale simulation of a Virtual Galaxy}\label{Sect:VGSim}

\begin{figure}
\begin{center}
  \psfig{figure=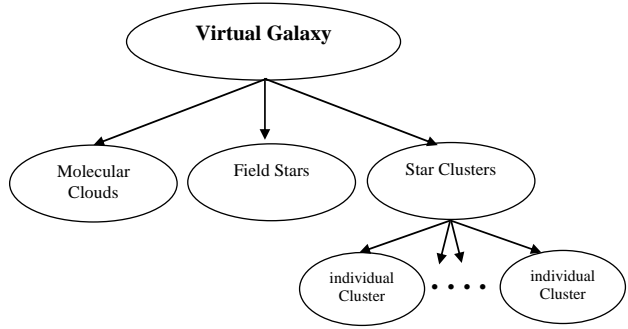,width=0.7\textwidth}
  \end{center}
  \caption{\label{fig:VGdecomp} Functional decomposition of the Virtual Galaxy}
\end{figure}

From Fig.\,\ref{fig:CPU} we see that the most expensive submodels in
a Virtual Galaxy are the star cluster simulations, the molecular
could simulations, and the field star simulations. In the following
discussion we neglect the other components. A Virtual Galaxy model,
viewed as a multi-scale multi-physics model, can then be decomposed
as in Fig.\,\ref{fig:VGdecomp}.

The by far most expensive operation is the star cluster
computations. We have $O(10^4)$ star clusters, each cluster can be
simulated independent of the others. This means that a further
decomposition is possible, down to the individual cluster level. A
single star cluster simulation, containing $O(10^4)$ stars, still
requires computational speeds at the TFlop/s scale (see also below).
The clusters simulations require $10^{21}$ Flop per simulated Myr of
lifetime of the Galaxy. The molecular clouds plus the field stars
need, on average over the full life time of the Galaxy, $10^{15}$
Flop per simulated Myr of lifetime, and can be executed on general
purpose parallel machines.

A distributed Petascale computing infrastructure for the Virtual
Galaxy could consist of a single or two general purpose parallel
machines to execute the molecular clouds and fields stars at a
sustained performance of 1 TFlop/s, and a distributed grid of
special purpose Grapes to simulate the star clusters. We envision
for instance 100 next generation GrapeDR systems\footnote{ Currently
some 100 Grape6 systems, delivering an average performance of 100
GFlop/s are deployed all over the world.}, each delivering 10
Tflop/s, providing a sustained 1 PFlop/s for the star cluster
computations. We can now estimate the expected runtime on a Virtual
Galaxy simulation on this infrastructure.  In Table\,\ref{tab:run}
we present the estimated wall-clock time needed for simulating the
Milky-way Galaxy, a smaller subset and a dwarf galaxy using the
distributed Petascale resource described above. Note that in the
reduced Galaxies the execution time goes linearly down with the
reduction factor, which should be understood as a reduction of mass
in the molecular clouds and a reduction of the total number of star
clusters (but with the same amount of stars per star cluster).

\begin{table}
\tabletitle{Estimated run times of the Virtual Galaxy simulation on
a distributed Petascale architecture as described in the main text.}
\label{tab:run}
\begin{tabular}{|c|c|c|c|}
  %\hline
  % after \\: \hline or \cline{col1-col2} \cline{col3-col4} ...
  Age & Milky Way Galaxy & Factor 10 reduction & Dwarf Galaxy \\
  & &  &(factor 100 reduction)\\
  \hline
  10 Myr & 3 hour & 17 min. & 2 min.  \\
  100 Myr & 3 year & 104 days & 10 days \\
  1 Gyr & 31 year & 3 year & 115 days \\
  10 Gyr & 320 year & 32 year & 3 year \\
  %\hline
\end{tabular}
\end{table}

With such a performance it will be possible to simulate the entire
Milky-way Galaxy for about 10\,Myr which is an interesting time
scale on which stars form, massive stars evolve and infant mortality
of young newly born star clusters operates. By simulating the entire
Milky-way Galaxy on this important time scale will enable us to
study these phenomena with unprecedented detail.

At the same performance it will be possible to simulate part
(1/10th) of the Galaxy on a time scale of 100\,Myr. This time scale
is important for the evolution of young and dense star clusters, the
major star formation mode in the Galaxy.

Simulating a dwarf galaxy, like the Large Magellanic Cloud for its
entire lifetime will become possible with a PFlop/s scale
distributed computer. The entire physiology of this galaxy is
largely not understood, as well as the intricate coupling between
stellar dynamics, gas dynamics, stellar evolution and dark matter.

\section{Discussion and Conclusions}\label{ch01sec05}
Multi-scale multi-science modeling is the next (grand) challenge in
Computational Science. Not only in terms of formulating the required
couplings across the scales or between multi-science models, but
also in terms of the sheer computational complexity of such models.
The later can easily result in requirements on the PFlop/s scale.

We have argued that simulating these models involves high level
functional decompositions, finally resulting in some collection of
single-scale single-science sub-models, that by themselves could be
quite large, requiring simulations on e.g. massively parallel
computers. In other words, the single-scale single-science
sub-models would typically involve some form of High Performance -
or High Throughput Computing. Moreover, they may have quite
different demands for compute infrastructure, ranging from
Supercomputers, via special purpose machines, to the single
workstation. We have illustrated this by pointing to a few models
from biomedicine and in more detail in the discussion on the Virtual
Galaxy.

We believe that the Grid provides the natural distributed computing
environment for such functionally decomposed models. The Grid has
reached a stage of maturity that in essence all the necessary
ingredients needed to develop a PFlop/s scale computational grid for
multi-scale multi-science simulations are available. Moreover, in a
number of projects grid enabled functionally decomposed distributed
computing has been successfully demonstrated, using many of the
tools that were discussed in Section \ref{ch01sec02}.

Despite these successes the experience with computational grids is
still relatively small. Therefore, a real challenge lies ahead in
actually demonstrating the feasibility of Grids for distributed
Petascale computing, and realizing Grid-enabled Problem Solving
Environments for multi-scale multi-science applications.

\end{document}